\documentclass[prl,letterpaper,twocolumn,showpacs,floatfix,superscriptaddress,amsmath,amsfonts,amssymb,preprintnumbers,citeautoscript]{revtex4}
\pdfoutput=1
\usepackage[T1]{fontenc}\usepackage[latin1]{inputenc}
\usepackage{dcolumn,graphicx,color,booktabs,microtype,afterpage} 
\usepackage[charter]{mathdesign}
\usepackage{sidecap}
\renewcommand{\figurename}{Fig.}
\renewcommand{\tablename}{Table}
\makeatletter\renewcommand{\fnum@figure}[1]{\figurename~\thefigure~(color online).}\makeatother
\makeatletter\renewcommand{\fnum@table}[1]{\tablename~\thetable.}\makeatother

\newcount\hh \newcount\mm
\hh=\time \divide\hh by 60
\mm=\hh \multiply\mm by 60 \mm=-\mm
\advance\mm by \time
\def\now{\number\hh:\ifnum\mm<10{}0\fi\number\mm}

\usepackage[colorlinks,plainpages=false,linkcolor=blue,urlcolor=blue,citecolor=blue,pdfpagemode=UseNone,pdfstartview=FitBH]{hyperref}

\newcommand{\half}{\frac{1}{\protect\raisebox{0.8pt}{\scriptsize 2}}}
\newcommand{\threehalf}{\frac{3}{\protect\raisebox{0.8pt}{\scriptsize 2}}}

\begin{document}

\makeatletter\renewcommand{\ps@plain}{%
\def\@evenhead{\hfill\itshape\rightmark}%
\def\@oddhead{\itshape\leftmark\hfill}%
\renewcommand{\@evenfoot}{\hfill\small{--~\thepage~--}\hfill}%
\renewcommand{\@oddfoot}{\hfill\small{--~\thepage~--}\hfill}%
}\makeatother\pagestyle{plain}


\title{Similar zone-center gaps in the low-energy spin-wave spectra of Na$_{1-\delta}$FeAs and BaFe$_2$As$_2$}

\author{J.\,T.~Park}\author{{\!\!$^{\!,}$\hyperref[FirstAuth]{\raisebox{1.5pt}{$^*$}}} G.\,Friemel}
\affiliation{Max-Planck-Institut für Festkörperforschung, Heisenbergstraße 1, 70569 Stuttgart, Germany}

\author{{\!\!$^{\!,}$\hyperref[FirstAuth]{\raisebox{1.5pt}{$^*$}}} T.~Loew}
\affiliation{Max-Planck-Institut für Festkörperforschung, Heisenbergstraße 1, 70569 Stuttgart, Germany}

\author{V.~Hinkov}
\affiliation{Quantum Matter Institute, University of British Columbia, Vancouver, B.C. V6T 1Z1, Canada}
\affiliation{Max-Planck-Institut für Festkörperforschung, Heisenbergstraße 1, 70569 Stuttgart, Germany}

\author{Yuan~Li}
\affiliation{Max-Planck-Institut für Festkörperforschung, Heisenbergstraße 1, 70569 Stuttgart, Germany}
\affiliation{International Center for Quantum Materials, School of Physics, Peking University, Beijing 100871, China}

\author{B.~H. Min}
\affiliation{Department of Emerging Materials Science, DGIST, Daegu 711-873, Republic of Korea}

\author{D.\,L.\,Sun}
\affiliation{Max-Planck-Institut für Festkörperforschung, Heisenbergstraße 1, 70569 Stuttgart, Germany}

\author{A.\,Ivanov}
\affiliation{Institut Laue-Langevin, 6 rue Jules Horowitz, 38042 Grenoble Cedex 9, France}

\author{A.\,Piovano}
\affiliation{Institut Laue-Langevin, 6 rue Jules Horowitz, 38042 Grenoble Cedex 9, France}

\author{C.\,T.~Lin}
\affiliation{Max-Planck-Institut für Festkörperforschung, Heisenbergstraße 1, 70569 Stuttgart, Germany}

\author{B.\,Keimer}
\affiliation{Max-Planck-Institut für Festkörperforschung, Heisenbergstraße 1, 70569 Stuttgart, Germany}

\author{Y.\,S.\,Kwon$^{4,\kern.5pt}$\hyperref[CorrAuthor1]{\raisebox{0.5pt}{$^\dagger$}}}
\noaffiliation

\author{D.\,S.\,Inosov$^{1,\,}$\hyperref[CorrAuthor2]{\raisebox{0.5pt}{$^\ddagger$}}}
\noaffiliation

\begin{abstract}
\noindent We report results of inelastic-neutron-scattering measurements of low-energy spin-wave excitations in two structurally distinct families of iron-pnictide parent compounds: Na$_{1-\delta}$FeAs and BaFe$_2$As$_2$. Despite their very different values of the ordered magnetic moment and N\'{e}el temperatures, $T_{\rm N}$, in the antiferromagnetic state both compounds exhibit similar spin gaps of the order of $10$\,meV at the magnetic Brillouin-zone center. The gap opens sharply below $T_{\rm N}$, with no signatures of a precursor gap at temperatures between the orthorhombic and magnetic phase transitions in Na$_{1-\delta}$FeAs. We also find a relatively weak dispersion of the spin-wave gap in BaFe$_2$As$_2$ along the out-of-plane momentum component, $q_z$. At the magnetic zone boundary ($q_z=0$), spin excitations in the ordered state persist down to $\sim$\,$20$\,meV\hspace{-0.5pt}, which implies a much smaller value of the effective out-of-plane exchange interaction, $J_{\rm c}$, as compared to previous estimates based on fitting the high-energy spin-wave dispersion to a Heisenberg-type model.
\end{abstract}

\pacs{75.30.Ds 74.70.Xa 78.70.Nx\vspace{-0.7em}}

\maketitle

The discovery of unconventional superconductivity in iron-pnictide compounds \cite{Discovery} with critical temperatures, $T_{\rm c}$, as high as 56\,K has fostered a tremendous interest in these materials in recent years \cite{Reviews}. There are several structurally distinct families of iron-based superconductors with similar phase diagrams \cite{Reviews}, governed by an interplay of antiferromagnetism, persistent in pure compounds under ambient pressure, and superconductivity that can be induced by pressure or chemical doping \cite{PaglioneGreene10}. Although the highest values of $T_{\rm c}$ are usually found in doped compounds with a nonstoichiometric composition, our physical understanding of these systems undoubtedly depends on the detailed knowledge of magnetic properties in the respective parent compounds, which are also easier to treat theoretically due to their simple crystal structure with no substitutional disorder.

Among the variety of such stoichiometric materials serving as parent phases for numerous iron-based superconductors, only a few have so far received proper experimental treatment, especially by inelastic neutron scattering (INS), due to miscellaneous reasons related to the availability of sizeable single crystals or their chemical stability. For instance, to the best of our knowledge, direct measurements of spin-wave excitations in the antiferromagnetic (AFM) state of iron pnictides have so far remained limited to a few members of the so-called `122' family with the ThCr$_2$Si$_2$-type structure, whose single crystals are typically stable in air and are readily available in the large sizes necessary for INS experiments. In particular, high-energy spin-wave modes have been mapped out in CaFe$_2$As$_2$ \cite{ZhaoAdroja09, DialloAntropov09} and BaFe$_2$As$_2$ \cite{HarrigerLuo11} using time-of-flight (TOF) neutron spectroscopy, which enabled estimations of the effective magnetic exchange interactions in the framework of a localized Heisenberg-type $J_{1a}$\,-\,$J_{1b}$\,-\,$J_{2}$\,-\,$J_{c}$ model. These results are complemented by INS measurements at lower energies, performed on polycrystalline BaFe$_2$As$_2$ \cite{EwingsPerring08} and on single crystals of SrFe$_2$As$_2$ \cite{ZhaoYao08}, CaFe$_2$As$_2$ \cite{McQueeneyDiallo08, DialloPratt10}, and BaFe$_2$As$_2$ \cite{MatanMorinaga09}. All of these measurements have unequivocally demonstrated the existence of a large anisotropy gap in the spin-wave dispersion at the magnetic Brillouin-zone center, which varies from 6.5\,--\,7.0\,meV in SrFe$_2$As$_2$ and CaFe$_2$As$_2$ \cite{ZhaoYao08, McQueeneyDiallo08, DialloPratt10} to almost 10\,meV in BaFe$_2$As$_2$ \cite{EwingsPerring08, MatanMorinaga09}. A more recent polarized INS experiment has revealed two distinct components of this gap in BaFe$_2$As$_2$, characterized by the out-of-plane and in-plane polarizations \cite{QureshiSteffens12}, with the onsets at 10\,meV and 16\,meV\/, respectively.

At present, first-principles calculations are faced with apparent difficulties in reproducing these energy scales in the spin-wave spectra even on a qualitative level \cite{YareskoLiu09, QureshiSteffens12}. Moreover, the identical crystal structure of all measured compounds, distinct from the structures of other families, precludes generalizations to all iron pnictides, making it difficult to relate the measured gaps to such microscopic structural and magnetic properties of the material as the ordered magnetic moment, exchange interactions, or crystallographic parameters. Therefore, in the present study we have performed INS measurements of the low-energy spin-wave spectrum in Na$_{1-\delta}$FeAs \cite{SasmalLv09} with the `111'-type structure, which we compare to that of BaFe$_2$As$_2$.

A single crystal of Na$_{1-\delta}$FeAs with a small Na deficiency, $\delta\approx0.1$, as estimated by energy-dispersive x-ray analysis, and a mass of $\sim$\,0.5\,g was grown by the vertical Bridgman method \cite{SongGhim10}. Pure FeAs$_{1.17}$ precursors were first synthesized from the reaction of high-purity Fe (powder, 99.999\%) and As (chips, 99.999\%) in sealed quartz containers at 1050$^\circ$C. A Na$_{1-\delta}$FeAs single crystal was then grown from a mixture of Na and precursor FeAs$_{1.17}$ with a molar ratio of 2:1 in a sealed molybdenum crucible. Higher molar concentrations of Na and As were necessary because of their high vapor pressures. During growth, the center of the furnace was heated to 1450$^\circ$C. Because of the extreme chemical sensitivity of Na$_{1-\delta}$FeAs to oxygen and air moisture \cite{TodorovChung10}, meticulous care had to be taken to exclude any contact with air while handling the crystal over the entire process of sample preparation and measurements. Prior to the INS experiment, the crystal had been sealed inside an aluminum can in helium atmosphere and oriented using a 4-circle neutron diffractometer \textit{Morpheus} of the Paul Scherrer Institute (PSI), Switzerland. These preliminary measurements indicated that the sample consists predominantly of one single-crystalline grain with a mosaicity $<0.5^\circ$. Our second sample was a mosaic of BaFe$_2$As$_2$ single crystals with a total mass of $\sim$\,0.9\,g, grown by the self-flux method as described elsewhere \cite{BFA_growth}, coaligned on a silicon wafer using x-ray Laue diffraction to a mosaicity $\lesssim$\,1.0$^\circ$.\enlargethispage{3pt}

\begin{figure}[t]\vspace{-5pt}
\includegraphics[width=\columnwidth]{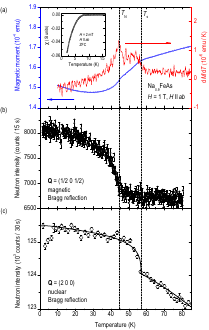}
\caption{Determination of the structural and AFM transition temperatures in Na$_{0.9}$FeAs from (a)~magnetization measurements, (b)~magnetic Bragg intensity, and (c)~nuclear Bragg intensity. Magnetic susceptibility data in the inset of panel (a) show the SC transition.}
\label{Fig:Tn_Ts}
\vspace{-13pt}
\end{figure}

A remarkable property of the Na$_{1-\delta}$FeAs compounds is the large splitting in temperature between the magnetic and structural phase transitions, which is present even in the parent phase \cite{ChenHu09, LiCruz09, ParkerSmith10, WrightLancaster12}, whereas in the `122' family of iron pnictides the two transition temperatures usually merge together upon the reduction of doping \cite{HuangQiu08, LesterChu09, NandiKim10, PrattTian09}. This makes it possible to study the narrow temperature window between the two transitions, which is typically associated with a so-called ``electronic liquid crystal'' (or ``spin nematic'') phase with a spontaneously broken rotational symmetry of the electronic eigenstates. In our Na$_{1-\delta}$FeAs sample, the AFM and orthorhombic phase transitions appear as anomalies in the temperature derivative of the magnetization [Fig.\,\ref{Fig:Tn_Ts}\,(a)] and in the order-parameter-like dependencies of the magnetic and nuclear Bragg peak intensities [Fig.\,\ref{Fig:Tn_Ts}\,(b,\,c)], respectively. From these measurements, the corresponding transition temperatures, $T_{\rm N}=45$\,K and $T_{\rm s}=57$\,K, could be determined, which are in agreement with literature values for samples of similar composition \cite{ChenHu09, LiCruz09, ParkerSmith10, WrightLancaster12}. Although the orthorhombic distortion is too weak to be directly resolved as the splitting of Bragg reflections in our experiment, the abrupt change in the (200) nuclear Bragg intensity at $T_{\rm s}$, seen in Fig.\,\ref{Fig:Tn_Ts}\,(c), is explained by the extinction release associated with a minor change in the sample's mosaicity across the orthorhombic transition, and has been previously used by several authors as a convenient and highly sensitive probe of the weak structural distortion \cite{LesterChu09, PrattTian09, Extinction}. In addition, our Na$_{1-\delta}$FeAs sample exhibits a weak superconducting (SC) transition at $T_{\rm c,\,onset}\approx8$\,K, but the volume fraction of the SC phase is below 10\%, according to magnetization measurements [Fig.\,\ref{Fig:Tn_Ts}\,(a), inset], and can be therefore neglected for the purpose of the present study. In BaFe$_2$As$_2$, both neutron diffraction and magnetization measurements (not shown) have revealed the AFM ordering below $T_{\rm N}=137$\,K, in agreement with previous reports \cite{HuangQiu08}.

\begin{figure}[t]\vspace{-1.7ex}
\includegraphics[width=\columnwidth]{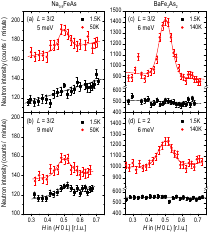}
\caption{Typical unprocessed constant-energy scans for Na$_{0.9}$FeAs (left) and BaFe$_2$As$_2$ (right), measured above and below $T_{\rm N}$ along the $|\mathbf{Q}|=\text{const}$ trajectories in momentum space, centered at $(\half 0 L)$. The values of $L$ at the center of the scan and the energy transfer are indicated in every panel.}
\label{Fig:Qscans}
\vspace{-10pt}
\end{figure}

Unprocessed INS data shown in Fig.\,\ref{Fig:Qscans} illustrate several representative momentum scans along the $|\mathbf{Q}|=\text{const}$ trajectories in the $(H0L)$ plane, measured on both samples above and below $T_{\rm N}$ in the spin-gap region. Here and henceforth, we are using the unfolded reciprocal-space notation of the Fe sublattice because of its simplicity and the natural correspondence to the symmetry of the observed signal \cite{ParkInosov10}. The wave vector $\mathbf{Q}=(HKL)$ is given in reciprocal-lattice units (r.l.u.), i.e.~in units of the reciprocal-lattice vectors of the Fe sublattice, in which the AFM ordering vector is $\mathbf{Q}_{\rm AFM}=(\half\,0\,\half)$. These measurements were done with a fixed final neutron momentum, $k_{\rm f}=2.662$\,\AA$^{-1}$, and with a pyrolytic graphite filter to eliminate higher-order reflections from the analyzer. In the paramagnetic state (Fig.\,\ref{Fig:Qscans}, circles), a commensurate peak is observed at the ordering vector down to the lowest energies, indicating the presence of gapless paramagnon excitations. Below the AFM ordering temperature, its intensity vanishes in the low-energy region, below the spin gap energy [panels (a) and (c)], or gets partially suppressed at intermediate energies close to the gap onset [panel (b)]. For BaFe$_2$As$_2$, we also show the corresponding scan centered at $L=2$ [at the magnetic zone boundary, panel (d)], where the situation is similar with the exception of a reduced intensity of the signal. The presence of a relatively strong paramagnetic intensity on the $(\half\,0\,L)$ line even far away from the ordering wave vector has been already pointed out in earlier works \cite{MatanMorinaga09, DialloPratt10}. It can be well understood in terms of Fermi surface nesting, which is maximized near $\mathbf{Q}_{\rm AFM}$, but still remains substantial for all values of $L$, according to band structure calculations \cite{ParkInosov10}.

\begin{figure}[t]\vspace{-6pt}
\includegraphics[width=\columnwidth]{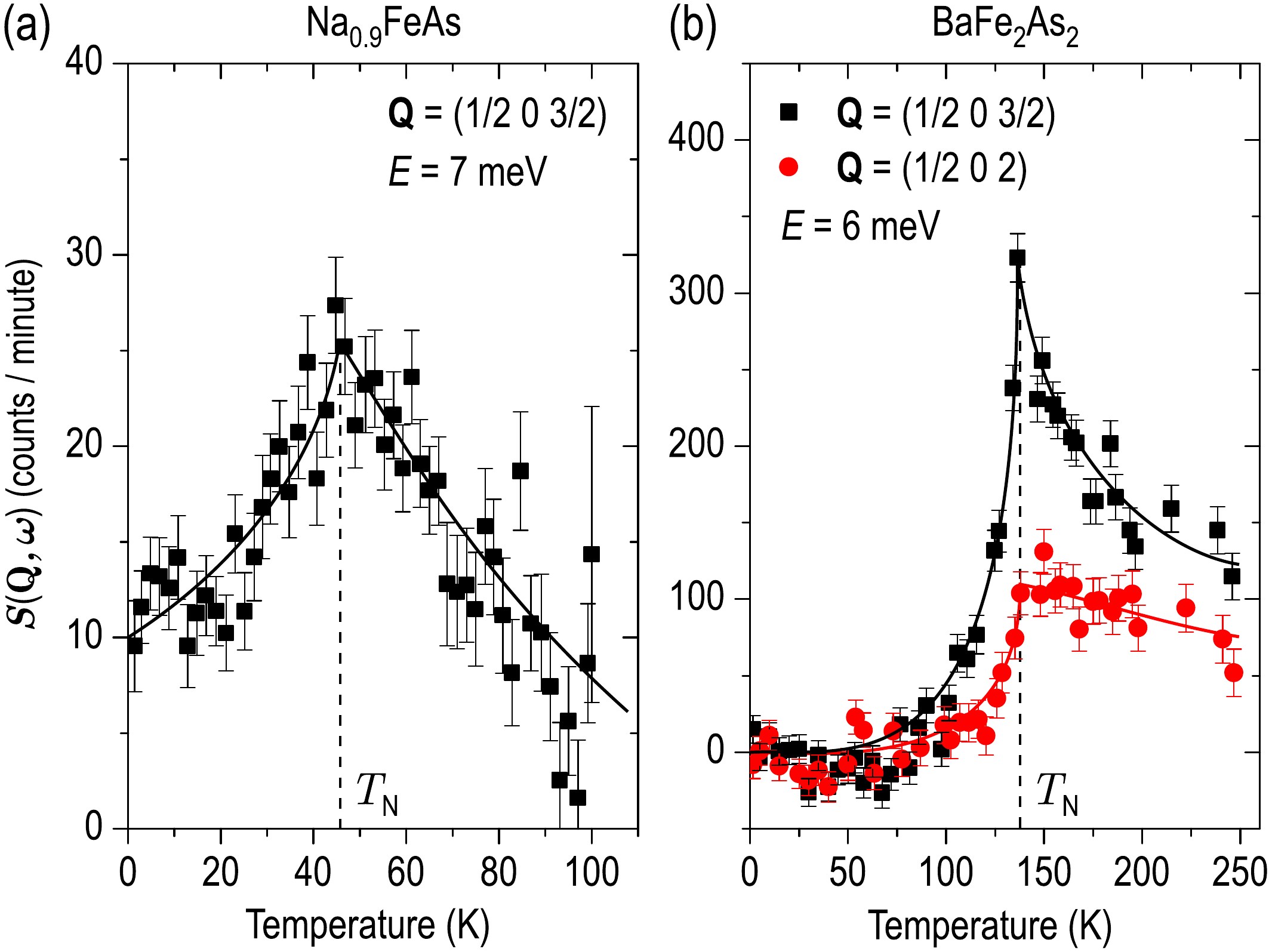}
\caption{Temperature dependence of the background-subtracted low-energy INS signal in Na$_{0.9}$FeAs (left) and BaFe$_2$As$_2$ (right), indicating the abrupt spin-gap opening below $T_{\rm N}$. Solid lines are guides to the eyes. Note the absence of any pronounced anomaly at the structural phase transition in Na$_{0.9}$FeAs ($T_{\rm s}=57$\,K).}
\label{Fig:Tdep}
\vspace{-13pt}
\end{figure}

The detailed temperature dependence of the low-energy INS signals in both samples is illustrated by Fig.\,\ref{Fig:Tdep}. Again, one can appreciate the similarity between the magnetic response measured at $\mathbf{Q}_{\rm AFM}$ (half-integer $L$) and at the magnetic zone boundary (integer $L$) in BaFe$_2$As$_2$ at an energy transfer of 6\,meV [Fig.\,\ref{Fig:Tdep}\,(b)]. Despite the somewhat lower amplitude of the signal at $(\half\,0\,2)$ as compared to $(\half\,0\,\threehalf)$, both increase monotonically upon cooling towards $T_{\rm N}$, where they exhibit a sharp kink due to the onset of the spin gap in the AFM ordered state. This anomaly is much sharper at $\mathbf{Q}_{\rm AFM}$ due to the critical scattering intensity at the ordering wave vector around $T_{\rm N}$.

It is remarkable that despite the presence of two well separated phase transitions in Na$_{1-\delta}$FeAs, the low-energy magnetic spectrum is only sensitive to the AFM transition and exhibits no pronounced anomaly at $T_{\rm s}$ outside of our experimental uncertainty [Fig.\,\ref{Fig:Tdep}\,(a)]. It is interesting to consider this observation in the framework of various models describing the electronic ``nematic'' state that was claimed responsible for the orthorhombic distortion in the narrow temperature range $T_{\rm N} < T < T_{\rm s}$ \cite{NematicModels}. Since a comprehensive theory of such a ``nematic'' electronic state in iron arsenides still awaits development, no definitive predictions about its magnetic excitation spectrum have been proposed. It has been recently suggested, however, that with the onset of a preemptive nematic order, the magnetic correlation length should discontinuously increase below $T_{\rm s}$, leading to a possible spectral weight redistribution and a consequent formation of a pseudogap \cite{FernandesChubukov12}. This could possibly result in a partial precursor gapping of the low-energy spin excitations already below $T_{\rm s}$, which is not confirmed by our measurements.

\begin{figure}[t]\vspace{-6pt}
\hspace*{-1.1em}\includegraphics[width=1.05\columnwidth]{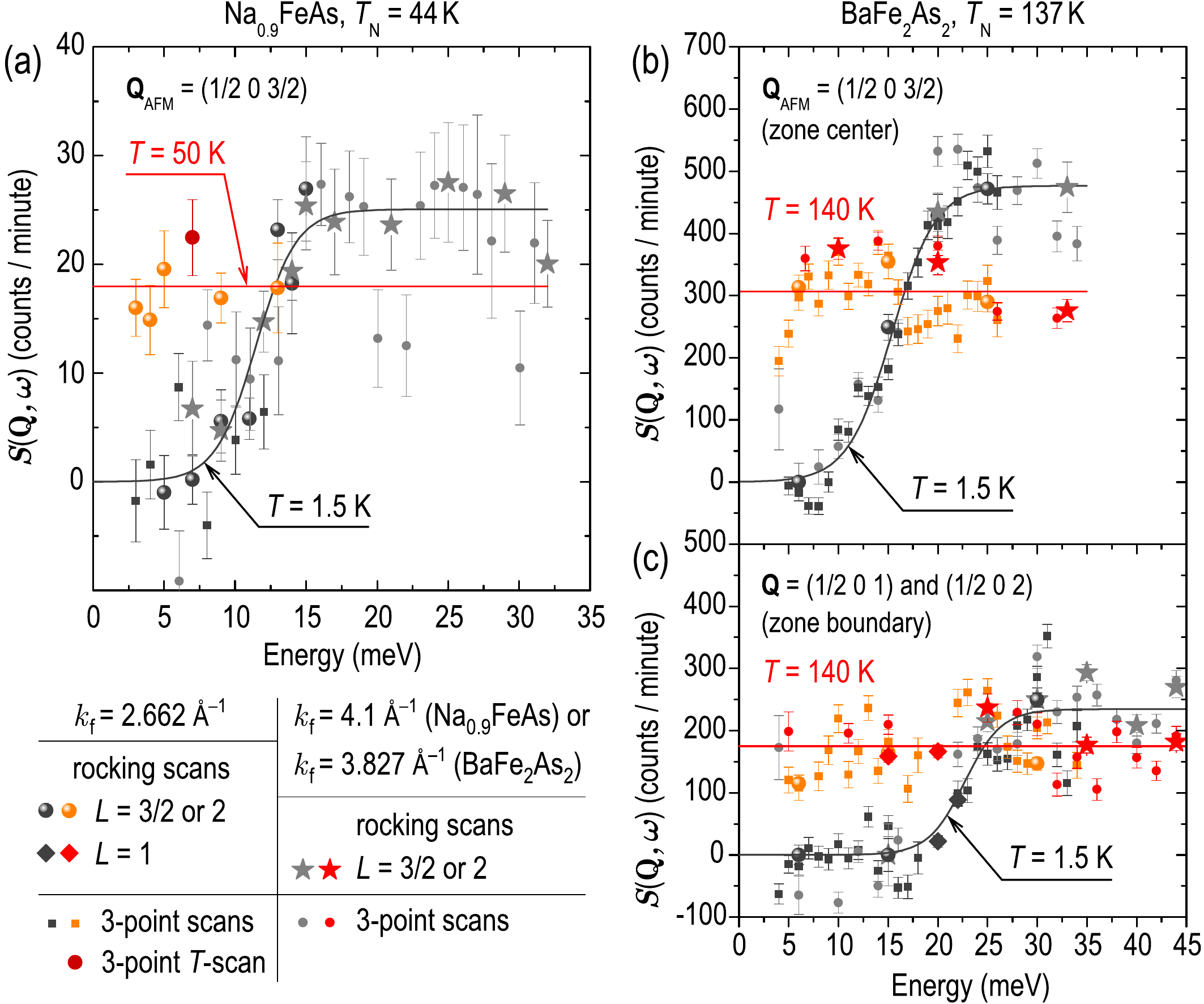}
\caption{Energy dependence of the background-subtracted magnetic INS intensity for Na$_{0.9}$FeAs (left) and BaFe$_2$As$_2$ (right), measured above and below $T_{\rm N}$, (a,\,b) at the ordering wave vector ($L=3/2$) and (c) at the magnetic zone boundary ($L=1,\,2$).}
\label{Fig:Gaps}
\vspace{-10pt}
\end{figure}

In Fig.\,\ref{Fig:Gaps}, we plot the energy dependence of the back\-ground-subtracted magnetic intensity, obtained from Gaussian fits of the momentum scans similar to those shown in Fig.\,\ref{Fig:Qscans} (larger symbols) or from 3-point measurements, in which the background intensity was obtained at two points on both sides of the peak (smaller symbols). Measurements with $k_{\rm f}=2.662$\,\AA$^{-1}$, $k_{\rm f}=3.837$\,\AA$^{-1}$ and $k_{\rm f}=4.1$\,\AA$^{-1}$ are shown with different symbols. In the paramagnetic state, a gapless spectrum of spin fluctuations with nearly energy-independent intensity is observed both at integer and half-integer $L$. As the temperature is decreased below the AFM transition, the low-energy spectral weight is transferred to higher energies, resulting in a clear spin gap in the magnetic excitation spectrum (see also Fig.\,\ref{Fig:Tdep}). At $\mathbf{Q}_{\rm AFM}$ (half-integer $L$), the onset of magnetic intensity at $T=1.5$\,K is observed at approximately $\sim$\,10\,meV both in the Na$_{0.9}$FeAs and BaFe$_2$As$_2$ compounds, so that the low-temperature spectra in Figs.~\ref{Fig:Gaps}\,(a) and (b) are nearly indistinguishable within the experimental uncertainty. Based on the recent polarized INS measurements \cite{QureshiSteffens12}, we can ascribe this onset to the smaller out-of-plane anisotropy gap. The onset of the in-plane scattering that occurs at a slightly higher energy can not be resolved as a separate step in our unpolarized data.

At the magnetic zone boundary (integer $L$), the gap in BaFe$_2$As$_2$ is only twice larger than at the zone center and amounts to approximately 20\,meV\hspace{-0.5pt}, in agreement with the assumptions of Ref.~\onlinecite{HarrigerSchneidewind09}. This clearly refutes the commonly accepted viewpoint that zone-boundary spin waves in BaFe$_2$As$_2$ are limited to high energies of the order of 50\,--\,100\,meV \cite{HarrigerLuo11, QureshiSteffens12}. In terms of the Heisenberg-type exchange interaction constants, our result suggests a much weaker out-of-plane exchange coupling, $J_{\rm c}$, than previously estimated from fitting high-energy TOF data to the anisotropic Heisenberg model \cite{HarrigerLuo11}. Indeed, assuming the in-plane exchange interactions ($J_{\rm 1a}$, $J_{\rm 1b}$, and $J_{\rm 2}$) of Ref.\,\onlinecite{HarrigerLuo11} to be unchanged, we can use the zone-center and zone-boundary gap magnitudes from our present study (10 and 20 meV\hspace{-0.5pt}, respectively) to re-estimate the two remaining parameters in the model: the effective out-of-plane exchange energy, $SJ_{\rm c}=0.22$\,meV\hspace{-0.5pt}, and the single ion anisotropy constant, $SJ_{\rm s}=0.14$\,meV\hspace{-0.5pt}. This reevaluated value of $J_{\rm c}$ is almost one order of magnitude smaller than previously reported \cite{HarrigerLuo11}, leading to a better agreement with the spin-wave velocities estimated from the nuclear-magnetic-resonance data \cite{NMR_BFA}.

\begin{table}[b]
\begin{center}
\begin{tabular}{l@{\hspace{-1em}}r@{~~~}l@{~~~}l@{\hspace{-1.4em}}r}\toprule
Compound & $T_{\rm N}$ (K) & $\mu_{\rm Fe}/\mu_{\rm B}$ & $\Delta_{\mathbf{Q}_{\rm AFM}}$ (meV) & Ref.\\
\midrule
SrFe$_2$As$_2$      & 205 & 1.0 \cite{KanekoHoser08+}            & ~~6.5 & \cite{ZhaoYao08}\\
CaFe$_2$As$_2$      & 172 & 0.8 \cite{GoldmanArgiryou08}         & ~~6.9(2) & \cite{McQueeneyDiallo08, DialloPratt10}\\
BaFe$_2$As$_2$      & 137 & 0.9 \cite{HuangQiu08,WilsonYamani09} & ~~7.7(2) & \cite{EwingsPerring08}\\
                    &     &                                      & ~~9.8(4) & \cite{MatanMorinaga09}\\
                    &     &                                      & 10.1 (out-of-plane) & \cite{QureshiSteffens12}\\
                    &     &                                      & 16.4 (in-plane) & \cite{QureshiSteffens12}\\
                    &     &                                      & 10(1) & [our work]\\
Na$_{1-\delta}$FeAs & 45  & 0.1 \cite{LiCruz09, WrightLancaster12}& 10(2) & [our work]\\
BaFe$_{1.96}$Ni$_{0.04}$As$_2$ & 91 &                            & $\!\!\!\sim$\,2 & \cite{HarrigerSchneidewind09}\\
\bottomrule
\end{tabular}\vspace{-8pt}
\end{center}
\caption{Comparison of the N\'{e}el temperatures ($T_{\rm N}$), values of the ordered magnetic moment ($\mu_{\rm Fe}$), and zone-center gap energies ($\Delta_{\mathbf{Q}_{\rm AFM}}$) in several iron-arsenide compounds.}
\label{Tab:Gaps}
\vspace{-10pt}
\end{table}

In Table~\ref{Tab:Gaps}, we have summarized several parameters of the AFM state, such as the N\'{e}el temperature ($T_{\rm N}$), value of the ordered magnetic moment ($\mu_{\rm Fe}$), and zone-center gap energy ($\Delta_{\mathbf{Q}_{\rm AFM}}$) for various iron-arsenide compounds. First, we observe that despite the tenfold difference in the ordered magnetic moment and a much lower ordering temperature, Na$_{1-\delta}$FeAs exhibits an anisotropy gap of the same order of magnitude as the materials of the `122' family. This is highly unusual, as theory predicts the anisotropy gap in spin-density-wave compounds to increase monotonically with the value of the sublattice magnetization, following a simple power law \cite{FishmanLiu98}, whereas our observations point to an anomalous behavior of the magneto-crystalline anisotropy in iron pnictides that is inconsistent with this general trend.

Second, we note that in `122'-compounds the spin gap is rapidly reduced upon doping Ni or Co into the Fe plane. As a result, in the doping range, where superconductivity coexists with static AFM order, spin fluctuations can already be observed below $T_{\rm N}$ at energies as low as 2--3\,meV \cite{HarrigerSchneidewind09, ChristiansonLumsden09}. Whenever these fluctuations extend below the energy of the SC gap, $2\Delta$, they can possibly serve as a source of spectral weight for the formation of a magnetic resonant mode below $T_{\rm c}$, which has been reported even in strongly underdoped `122' samples with $T_{\rm c}$ as low as 11\,K \cite{ChristiansonLumsden09}. In contrast to this scenario, superconductivity in the phase diagram of doped NaFeAs is found in the immediate vicinity of the parent phase \cite{ParkerSmith10, WrightLancaster12}, where we have shown the anisotropy gap to be much larger than $2\Delta$. Indeed, for our sample with $T_{\rm c}=8$\,K, even the exceptionally high ratio of $2\Delta/k_{\rm B}T_{\rm c} \approx 8$ reported by Liu \textit{et al.} \cite{LiuRichard11} would result in $2\Delta$ of only 5.5\,meV, whereas the weak-coupling ratio of $2\Delta/k_{\rm B}T_{\rm c} = 3.53$ yields $2\Delta = 2.4$\,meV, which is 4 times smaller than the magnetic anisotropy gap. Therefore, if the conventional scenario for the formation of the magnetic resonant mode also holds in the Na-111 family of superconductors, only 1--2\% of Co or Ni would have to induce a substantial rearrangement of the low-energy magnetic spectral weight in NaFeAs, in order to reduce the zone-center gap and lead to a finite magnetic intensity below $2\Delta$. Further INS experiments on doped samples are required to explore the interplay between magnetism and superconductivity in this system.

The authors are grateful to G.\,Khaliullin, A.\,Yaresko, N.~Shannon, and A.~Smerald for fruitful discussions and thank J.~White and C.~Busch for the assistance in sample characterization. All presented INS data were collected using the triple-axis thermal-neutron spectrometer IN8 at the Institut Laue-Langevin (ILL) in Grenoble. This work has been supported, in part, by the DFG within the Schwerpunktprogramm 1458, under Grant No. BO3537/1-1, and by the MPI\,--\,UBC Center for Quantum Materials. B.\,H.\,M. and Y.\hspace{2.5pt}S.\,K. acknowledge support from the Basic Science Research Program (Grant No. 2010-0007487) and the Mid-career Researcher Program (Grant No. 2010-0007487) through NRF funded by the Ministry of Education, Science and Technology of Korea. Y.\,L. acknowledges support from the Alexander von Humboldt Foundation.

\end{document}